\documentclass[aps,prl,twocolumn,amsmath,notitlepage,floatfix,footinbib,superscriptaddress,showpacs, showkeys]{revtex4-1}
\usepackage{graphicx}
\usepackage{amssymb}
\usepackage[utf8]{inputenc}
\usepackage[usenames,dvipsnames]{color}
\usepackage[disable]{todonotes}
\usepackage{hyperref}

\newcommand{\Tr}{\text{Tr}}
\renewcommand{\Im}{\mathrm{Im}}
\renewcommand{\Re}{\mathrm{Re}}

\begin{document}
\title{Voltage quench dynamics of a Kondo system} 

\author{Andrey~E.~Antipov}\email{aantipov@umich.edu}
\author{Qiaoyuan Dong}
\author{Emanuel Gull}
\affiliation{Department of Physics, University of Michigan, Ann Arbor, Michigan 48109, USA}

\date{\today}

\begin{abstract}
We examine the dynamics of a correlated quantum dot in the mixed valence regime. We perform numerically exact calculations of the current after a quantum quench from equilibrium by rapidly applying a bias voltage in a wide range of initial temperatures. The current exhibits short equilibration times and saturates upon the decrease of temperature at all times, indicating Kondo behavior both in the transient regime and in steady state. The time-dependent current saturation temperature matches the Kondo temperature at small times or small voltages; a substantially increased value is observed outside of linear response. These signatures are directly observable by experiments in the time-domain.
\end{abstract}



\maketitle

\listoftodos 

The Kondo effect is an intrinsically many-body phenomenon in which localized and itinerant electrons form a strongly correlated state \cite{Kondo1964,Abrikosov1965,Suhl1965} which shows signatures in thermodynamic, spectral, and linear response transport properties \cite{Costi1994,Konik2001}. Originally introduced to describe the low-temperature properties of magnetic impurities embedded in a non-magnetic bulk \cite{Kondo1964}, it has since been observed in a wide variety of correlated electron systems ranging from impurities adsorbed on surfaces \cite{Madhavan1998,Knorr2002} to molecular transistors \cite{Park2002}, heavy fermion materials \cite{Gegenwart2008}, and mesoscopic quantum wires \cite{Schopfer2003}. 
Kondo behavior is also observed in semiconducting quantum dot heterostructures, where a confined interacting region is coupled to non-interacting leads \cite{Goldhaber-Gordon1998,Kouwenhoven2001,Jeong2001,DeFranceschi2002}. Gating of these systems allows to study Kondo physics over a wide parameter range\cite{Goldhaber-Gordon1998}. 

Kondo correlations only emerge at low temperature. Their onset is characterized by the Kondo temperature $T_K$ that can be defined as the temperature at which the zero-bias conductance reaches half of its low-temperature value \cite{Costi1994}.
In thermal equilibrium, the Kondo problem is well understood and quantitatively described by a range of analytical and numerical methods \cite{Wilson1975,Hewson:1993,Bulla2008,Schollwock2005,Hirsch1986,Gull2011a,Rubtsov2005,Werner2006}.

When a Kondo system is driven out of equilibrium, additional phenomena appear. For instance, the application of a (time-independent) bias voltage splits the Kondo peak \cite{Meir1992b, DeFranceschi2002, Fujii2003, Plihal2005,Han2007,Fritsch2010,Dirks2010,Mitra2011, Dorda2014,Cohen2014a} and shows signatures in the double occupancy and magnetization \cite{Dirks2013}. At large voltages the voltage dependence of the conductance decreases on an energy scale comparable to $T_K$ \cite{Rosch2005,Kaminski2000, Kretinin2012, Smirnov2013} and its temperature dependence saturates at temperatures above $T_K$ \cite{Wingreen1994,Kretinin2011, Reininghaus2014}. 

How observables evolve in time after a rapid change of parameters \cite{Latta2011,Tureci2011} and how they decay to their steady state limit is an open question. Recent experimental progress in the measurement of time-dependent quantities on ever faster time scales has enabled experimental studies of such transient dynamics \cite{Nunes1993,Loth2010,Terada2010,Yoshida2014}, making a theoretical description of quenches from correlated initial states important. 

In this paper we provide numerically exact results for the real time evolution of a Kondo system following a bias voltage quench from a correlated equilibrium ensemble to a non-equilibrium steady state. We show results for transient and steady state currents and populations at temperatures ranging from $T \gg T_K$ to $T \ll T_K$, which have not previously been accessible in numerical calculations. 
Our results, enabled by recent advances in numerically exact QMC methods, illustrate the time evolution of the current and its saturation temperature from the equilibrium Kondo temperature to the increased steady state value and predict experimentally observable signatures of many-body correlations far from linear response.

Quenches of initially uncorrelated systems have been  examined by a variety of state of the art theoretical methods. In such systems, electronic correlations are gradually established as time progresses. Studies by the iterative path integral approach (ISPI) \cite{Weiss2008,Segal2010,Eckel2010,Becker2012,Hutzen2012,Weiss2013}, real-time renormalization group \cite{Schoeller2009,Andergassen2011,Kennes2012}, hierarchical equations of motion (HEOM) \cite{Zheng2009, Hartle2014, Hartle2015}, flow-equation methods \cite{Wang2010}, exact solutions in solvable limits \cite{Heyl2010,Ratiani2010}, perturbation theory \cite{Hara2015}, time-dependent Gutzwiller approach \cite{Schiro2010b, Lanata2012}, DMRG \cite{White2004,Daley2004}, NRG \cite{Anders2006}, continuous time quantum Monte Carlo (QMC) \cite{Schmidt2008,Werner2009,Muhlbacher2011,Koga2013, Dirks2013} and bold-QMC \cite{Gull2010a,Gull2011,Cohen2014,Cohen2013a,Cohen2014a} have shown the dynamical build-up of Kondo correlations on exponentially long time scales, the time-dependent Kondo cloud formation in the leads \cite{Lechtenberg2014, Nuss2015}, and characterized in detail the steady state properties, including the current-voltage characteristics, voltage-split spectral functions and temperature dependence of the conductance. 

Quenches from a strongly correlated initial state pose a greater challenge, as the initial solution of an equilibrium many-body problem is required. At high temperature $T \gtrsim T_K$, the dynamics following the voltage quench has been described by QMC \cite{Werner2010, Schiro2010a} and HEOM \cite{Hartle2013, Ye2015, Hartle2015}. These simulations have demonstrated time-dependent currents and spectral functions and related current oscillations to the applied voltage \cite{Cheng2015}. At $T=0$ these results are complemented by DMRG \cite{Kirino2008, Heidrich-Meisner2009a, Kirino2011} and NRG \cite{Anders2006, Anders2008, Eidelstein2012} simulations, recently extended to finite temperature \cite{Nghiem2014}, which show transient dynamics between the ground and steady states.  Semi-analytic techniques, including NCA and OCA  \cite{Plihal2005, Goker2010, Oguri2013,Eckstein2010a,Aoki2014} and the time-dependent Gutzwiller approach \cite{Schiro2010b, Lanata2012} provide approximate results for all values of voltage and temperature. Their precision requires careful assessment \cite{Haule2001, Tosi2010}. In principle, QMC simulations can cover the full parameter range of the system. However, so far these calculations were limited to high temperature and short times by sign problems. We have overcome the first limitation in this work by successively normalizing to a sequence of reference systems at progressively higher expansion order, allowing us to gradually reach temperatures an order of magnitude below $T_K$. 

\begin{figure}[t]\begin{center}
\includegraphics[width=0.95\columnwidth]{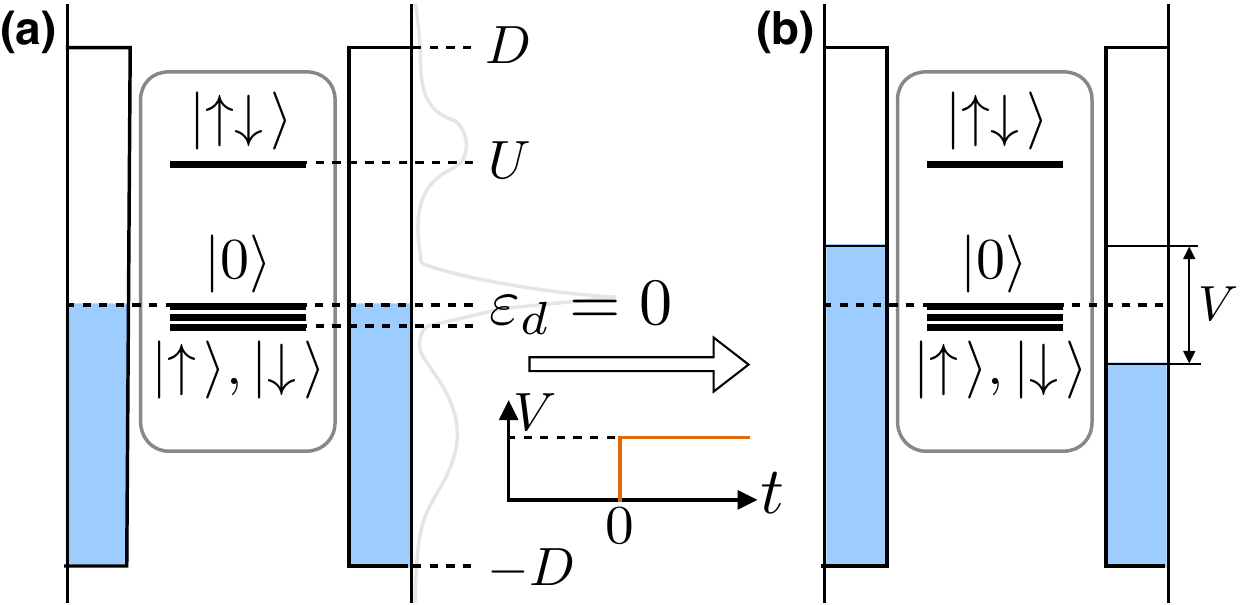}\end{center}\vspace{-2em}
\caption{(a) Interacting quantum dot embedded into a tunnel junction. The dot is described by a level spacing $\varepsilon_d$ and a charging energy $U$. The leads are described by the half-bandwidth $D$ and the chemical potential and $\varepsilon_d = 0$. (b) At time $t=0$ a voltage $V$ is instantaneously applied to the system, changing the bath Fermi level to $\pm V/2$. }\vspace{-1.5em}
\label{fig1}
\end{figure}

\emph{Model}.~We consider a quantum dot attached to two metallic leads, schematically shown in Fig.~\ref{fig1}. Both the quantum dot and the leads are initially in thermal equilibrium at temperature $T$. We describe the system by an Anderson impurity model 
\begin{subequations}
\begin{align}\label{eq:hamilt}
H(t)= & \sum_{\alpha k \sigma}(\varepsilon_{k} + \frac{\alpha}{2} V(t)) c_{\alpha k\sigma}^{\dagger}c_{\alpha k\sigma}+ H_T+H_D\\ H_T = & \mathcal{V}\sum_{k\sigma}(c_{\alpha k\sigma}^{\dagger}d_{\sigma}+d_{\sigma}^{\dagger}c_{\alpha k\sigma}) \\  H_D = &  \sum_{\sigma}\varepsilon_{d}d_{\sigma}^{\dagger}d_{\sigma}+Un_{\uparrow}n_{\downarrow}.
\end{align}
\end{subequations}
$c$ and $d$ label electrons in the leads and on the dot, respectively, $\sigma$ is the spin index, $\alpha = \pm 1$ labels the left $(+)$ and right $(-)$ lead. $\mathcal{V}$ is the tunneling matrix element between impurity and leads, $\varepsilon_k$ is the lead dispersion, $\varepsilon_d$ the impurity level spacing, and $U$ the impurity electronic repulsion strength (charging energy). The chemical potential is set to $0$ at equilibrium, the instantaneous (quenched) application of voltage $V(t) = V \theta(t) $ at time $t=0$ changes it to $\alpha \frac{V}{2}$. 

The leads are non-interacting and are described by a flat density of states $A(\omega)$ 
with half-bandwidth $D$ and a smooth cutoff at the band edges 
\cite{Werner2010}. The coupling of the dot and leads is described by the effective coupling parameter $\Gamma = \pi \mathcal{V}^2 / (2D)$ and the hybridization function 
\begin{multline}\label{eq:hyb_fun}
\Delta_\sigma(t,t') = \sum_\alpha \Delta_{\alpha\sigma}(t,t') = \sum_\alpha -i \mathcal{V}^2 \int_C d\omega A(\omega)\\  \times \left[\theta_{C}(t,t')-f(\omega)\right] \exp\left(-i\int_{t'}^{t}dt''(\omega - \alpha\frac{V(t'')}{2})\right).
\end{multline}
$\int_C$ denotes the integral over the L-shaped Keldysh contour with an imaginary branch added to take into account the initial correlated conditions \cite{Rammer1986}, and $f(\omega) = (\exp(\omega / T) +1)^{-1}$ is the Fermi function. 

We describe the evolution of the system by a hybridization expansion in the tunneling term $H_T$, which we stochastically sum to all orders using diagrammatic continuous-time Monte Carlo (CT-HYB) on the full Keldysh contour  \cite{Werner2006,Werner2009,Schiro2010a}. 
We perform an expansion for the partition function of the problem $Z [\Delta] = \sum_n \Tr_{|\Psi\rangle} \sum_{\phi_n} w_{\phi_n}$, where every perturbation order $n$ of expansion in $H_T$ is characterized by configurations $\phi_n$ consisting of $2n$ operators distributed on the contour and the impurity state $|\Psi\rangle$. 
Summing the series stochastically allows one to extract the time-dependence of the quantum dot charge \cite{Gull2011} as $N = \sum_{\phi} \langle \Psi |\hat N s_\phi |\Psi\rangle) / {\langle s \rangle}$, where $\sum_\phi = \sum_n \sum_{\phi_n}$, $s_\phi = \Re{w_{\phi_n}} / |\Re{w_{\phi_n}}|$ is the Monte Carlo sign of every configuration and $\langle s \rangle = \sum_{\phi} s_{\phi} $ is the average sign. The sign $\langle s \rangle$  is equal to $1$ in equilibrium, but decays exponentially with real time. In practice we resolve times of order of $\Gamma^{-1}$. Calculations at larger times are exponentially more expensive.

The main observable of interest is the current passing through the dot. The QMC procedure should be modified for current calculations by replacing $\Delta$ with $\Delta_t(t_1,t_2) = \Delta(t_1,t_2)(1 - \delta_{t_1 ,t}) + \Delta_{+}(t_1,t_2) \delta_{t_1,t}$ and sampling the imaginary part of the weight of every configuration, yielding $s^I_\phi = \Im{w_{\phi_n}} / |\Im{w_{\phi_n}}|$ \cite{Werner2009}. From this we extract the quantity $ZI(t) = 2 \Im Z[\Delta_t]$ and, by dividing by $Z$, the current.

In order to evaluate the current separate measurements of $Z$ and $ZI$ are required. These quantities are not directly accessible by QMC, but are instead inferred from a normalization procedure, yielding 
\begin{equation}\label{eq:current}
I(t) = \left[\frac{ZI(t)}{Z_{\mathrm{ref}} I_{\mathrm{ref}}(t)}\right] \left[ \frac{Z_{\mathrm{ref}}}{Z} \right] I_{\mathrm{ref}}(t), 
\end{equation}
where $Z_{\mathrm{ref}}$ and $I_{\mathrm{ref}}$ are the partition function and the current of a reference system, which are obtained separately. Each of the fractions in Eq. (\ref{eq:current}) is obtained in a CT-HYB calculation as
\begin{equation}
\frac{Z_\mathrm{ref}}{Z} =\frac{ \sum_{\phi} {\Re w_\mathrm{ref}} / |\Re w_{\phi}| }{ \sum_{\phi} {\Re w_\phi} / |\Re w_{\phi}|  }; \hspace{0.3em} \frac{ZI_\mathrm{ref}}{ZI} =\frac{ \sum_{\phi} {\Im w_\mathrm{ref}} / |\Im w_{\phi}| }{ \sum_{\phi} {\Im w_\phi} / |\Im w_{\phi}|  }
\end{equation}

The choice of the reference system is important: small ratios of $I_{\mathrm{ref}}/I$ and $Z_{\mathrm{ref}}/Z$ results in large relative error bars. As temperature is decreased, the average perturbation order increases $\propto T^{-1}$, and the normalization to the first order of expansion in $H_T$ \cite{Werner2009} becomes unreliable. We improve the algorithm in several ways: first, at high temperatures (perturbation orders $\leq 10$) the reference system is a finite order NCA/OCA expansion. At lower temperatures (larger pertubation orders) we perform a series of calculations, progressively increasing the perturbation order until convergence is achieved, yielding 
\begin{equation}
ZI = \left[\prod_{n=1}^{\infty} \frac{ZI_{m + n\delta}}{ZI_{m + (n-1)\delta}} \right] \left[\frac{ZI_{m}}{ZI_{\mathrm{NCA}}}\right] ZI_{NCA},
\end{equation}
with the order of the NCA expansion $m=7$ and $\delta$ as either $2$ or $3$ to maintain ratios $>10^{-1}$ at every step. 
We emphasize that our results are exact within the stochastic error bars for any choice of reference system; the choice affects only the computational cost. 

\emph{Results}.~We present results for the dynamics of a quantum dot in the mixed valence regime,  $\varepsilon_d = 0$. The Kondo temperature $T_K$ is on the order of the coupling $\Gamma$ \cite{Goldhaber-Gordon1998}, and observables are expected to equilibrate within times $\propto \Gamma^{-1}$, accessible by QMC calculations \cite{Werner2009,Gull2010a,Gull2011}.
For convenient comparison to experiment we set the unit of energy close to $1$ meV. Time is then given in meV${}^{-1} \approx 4$ ps. We parametrize the setup with  a coupling of $\Gamma = 0.3$, a charging energy $U = 3$ ($10 \Gamma$) and a half-bandwidth $D = 5$ ($15\Gamma$) similar to Ref. \onlinecite{Keller2013}. Band cutoff effects at $D > U$ are expected  not to play a role in the dynamics \cite{Werner2010,Hewson:1993}. 

\begin{figure}[ht!]
\includegraphics[width=\columnwidth]{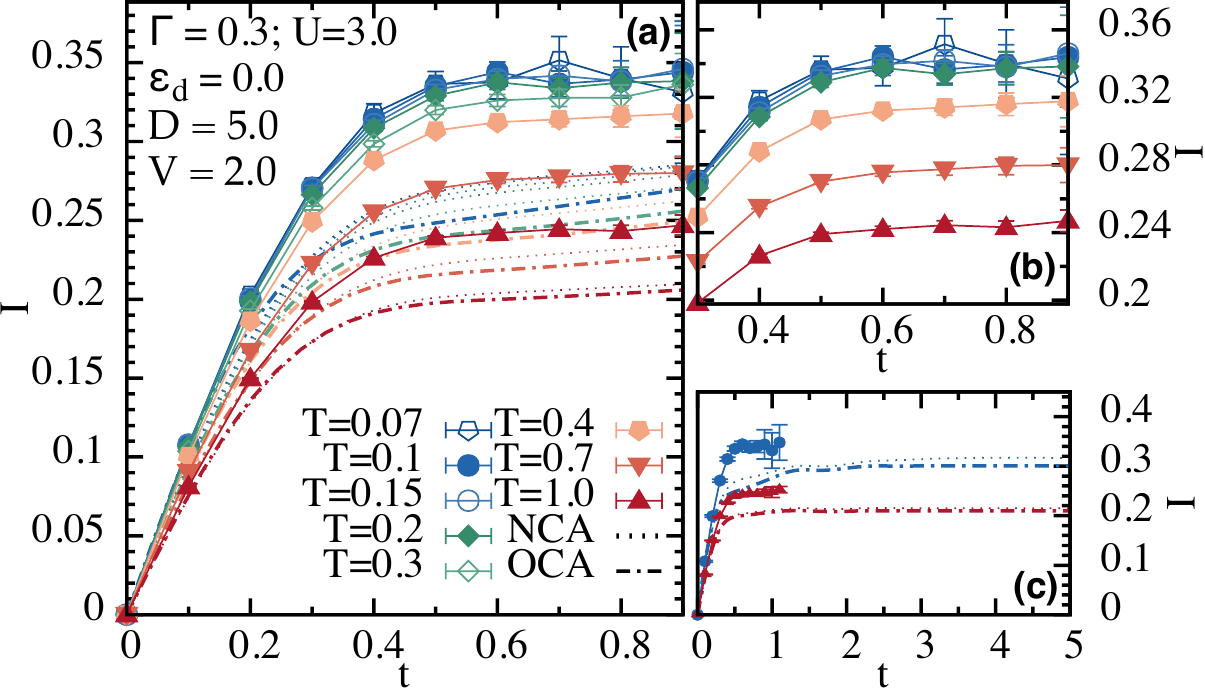}\vspace{-1.0em}
\caption{Current as a function of time $t$ after a voltage quench, for temperatures $T$ indicated. Parameters $U=3$, $\varepsilon_d = 0$, $\Gamma = 0.3$, $D=5$ and $V=2$. (a) Monte Carlo (solid symbols, full lines), NCA (dotted lines) and OCA (dash dotted lines) results. (b) Zoom at times $t=0.3 - 0.9$. (c) Behavior of the NCA and OCA currents up to times $t=5$.}
\label{fig2}
\end{figure}

Fig. \ref{fig2} shows the behavior of the current $I$ after a voltage quench. 
In panel (a) we show $I$ as a function of time $t$ following a voltage quench from $V = 0$ to $V = 2$ for a set of temperatures $0.07 \leq T \leq 1$ (between $0.7$ and $11$ K). 
Results from semi-analytical approximations are shown as fine dotted lines (NCA) and dash-dotted lines (OCA) for comparison \cite{Aoki2014,Eckstein2010a}. We observe that for all temperatures the current equilibrates at time $t \approx 0.6$ ($\approx 2.5$ ps) well within reachable times of $\sim 1$ ($4$ ps). As temperature is lowered from $T = 1$ ($11$ K) to $T \sim 0.2$ ($2$ K) the current at fixed times and in steady state increases. Further reducing $T$ by a factor of five yields no additional increase of current, illustrated by panel (b). 
We attribute the fast equilibration time to the high Kondo temperature of this mixed valence system \cite{Gull2011}.

Results from NCA and OCA correctly capture the short-time behavior and the overall shape of the current but underestimate both the transient and the steady state value by $\approx 20 \%$. This is known for systems with $U < D$ \cite{Haule2001}, although the quality of these methods will presumably improve in the strong interaction limit and by including vertex corrections in OCA \cite{Tosi2010}.  Both results are shown in panel (c) for times much longer than presently accessible by QMC. No additional time-dependence is visible, illustrating that our calculations are able to reach steady state.


\begin{figure}[ht!]
\includegraphics[width=\columnwidth]{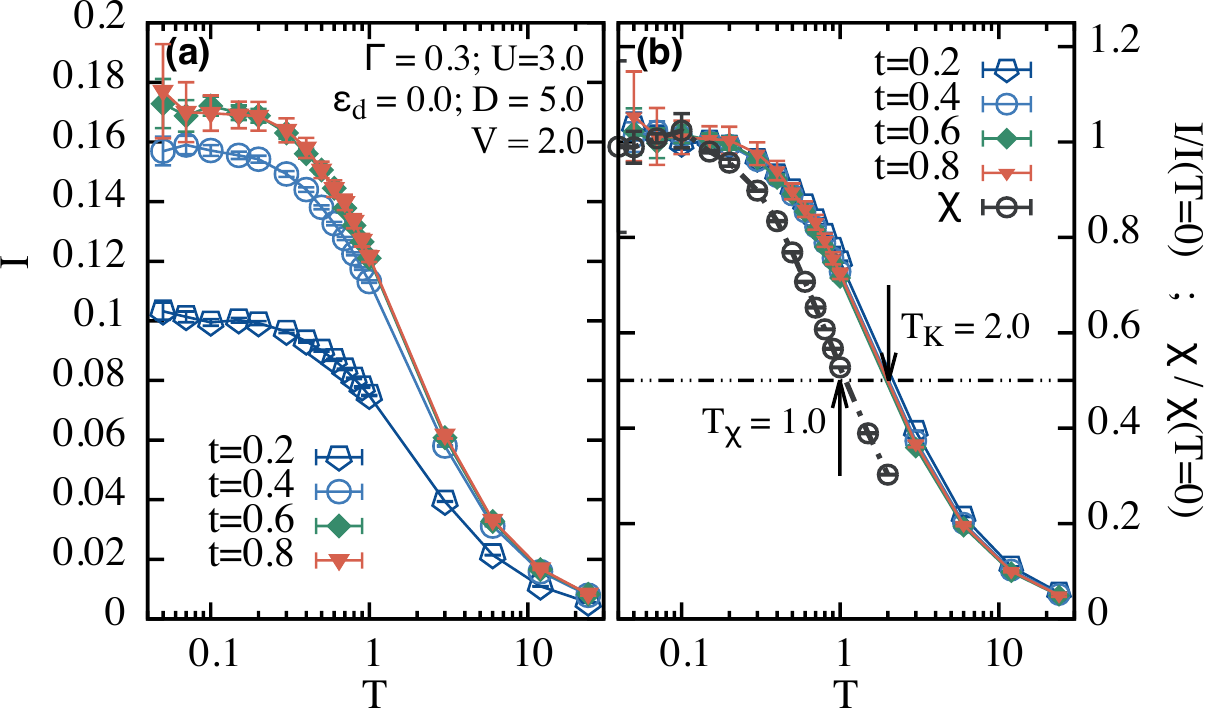}\vspace{-1.0em}
\caption{(a) Transient current after a voltage quench at times $t=0.2, 0.4, 0.6,0.8$ and $t=1.0$ as a function of temperature at $U=3, \varepsilon_d = 0, \Gamma = 0.3$, $D=5$ and voltage $V=2$. (b) Normalized current $I(T)/I(T=0)$ and equilibrium magnetic susceptibility $\chi(T)/\chi(T=0)$ as a function of temperature. Dashed line shows the value of $1/2$ and vertical arrows show the crossing points of this line. 
}
\label{fig3}
\end{figure}

The temperature dependence of the current is analyzed in Fig.~\ref{fig3}. The saturation of the low-$T$ current is clearly visible in panel (a), where different traces show the transient current obtained at different times. 
We identify the saturation of the current in the low-$T$ regime $T = 0.07 - 0.2$ with Kondo behavior and the temperature at which the current reaches half of its saturated low-$T$ value as the Kondo temperature $T_K$ \cite{Hewson:1993,Costi1994}. 

Fig.~\ref{fig3}b shows the temperature dependence of the current $I/I(0)$ and equilibrium magnetic susceptibility $\chi / \chi(0)$ normalized  to the respective zero-temperature values. The magnetic susceptibility, defined as a response to the infinitesimal local magnetic field $h$, $\chi = \left.\frac{d\langle n_{\uparrow} - n_{\downarrow}\rangle}{dh}\right|_{h\to 0}$ saturates at low temperatures \cite{Hewson:1993, Hanl2014}, similarly to the current. For $V=2$ all normalized current values for different times collapse on a single curve. This shows that Kondo behavior can be detected  based purely on short-time transient dynamics. The Kondo temperature $T_K$ determined by the current is time-independent and occurs at around twice the value of $T_\chi$ as defined from the point where the equilibrium magnetic susceptibility reaches half of its zero-$T$ value, as expected in a mixed-valence regime \cite{Goldhaber-Gordon1998, Yoshida2009,Smirnov2013}.

\begin{figure}[ht!]
\includegraphics[width=\columnwidth]{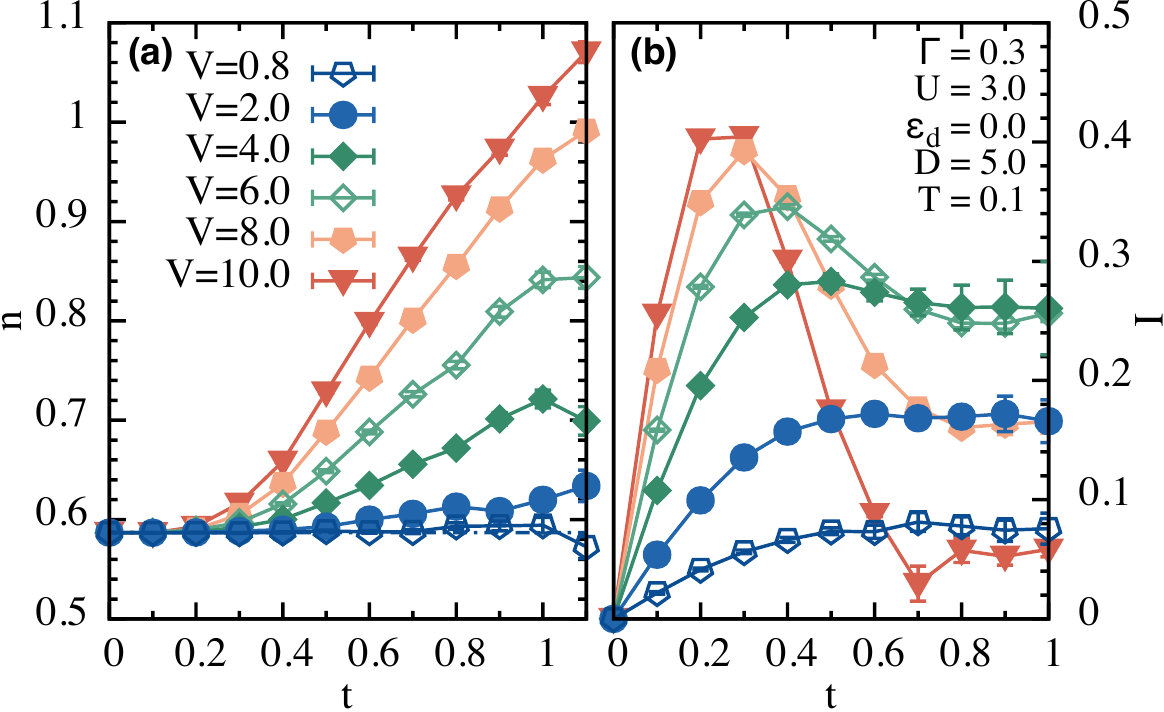}\vspace{-1.0em}
\caption{
(a) Quantum dot charge and (b) current at $U=3, \varepsilon_d = 0.0, \Gamma = 0.3$, $D=5.0$ and temperature $T=0.1$ for voltages $V = 0.8, 2, 4, 6$ (equal to $2U$), $8$, and $10$ (equal to $2D$).}
\label{fig4}
\end{figure}

We proceed with studying the impact of the magnitude of the applied bias voltage. Fig.~\ref{fig4} shows the time dependence of the quantum dot occupancy (a) and current (b) at a range of voltages between $0.8$ and $10$. At voltages $V\leq 2$ (linear response) the occupancy retains its equilibrium value and the current shows a monotonic rise and saturation to the steady state value, which increases with the applied voltage. Larger voltages $V > 2$ demonstrate linear response behavior only at small times $t < 0.3$. At larger times the nonlinear behavior is visible: the current decreases to the steady state value, which becomes smaller with applied $V$, illustrating the breakdown of conductance at large voltages \footnote{At $V \gg D$ the current decays to zero and oscillates with the period $\simeq 2\pi/V$ (not shown). }.

\begin{figure}[ht!]
\includegraphics[width=\columnwidth]{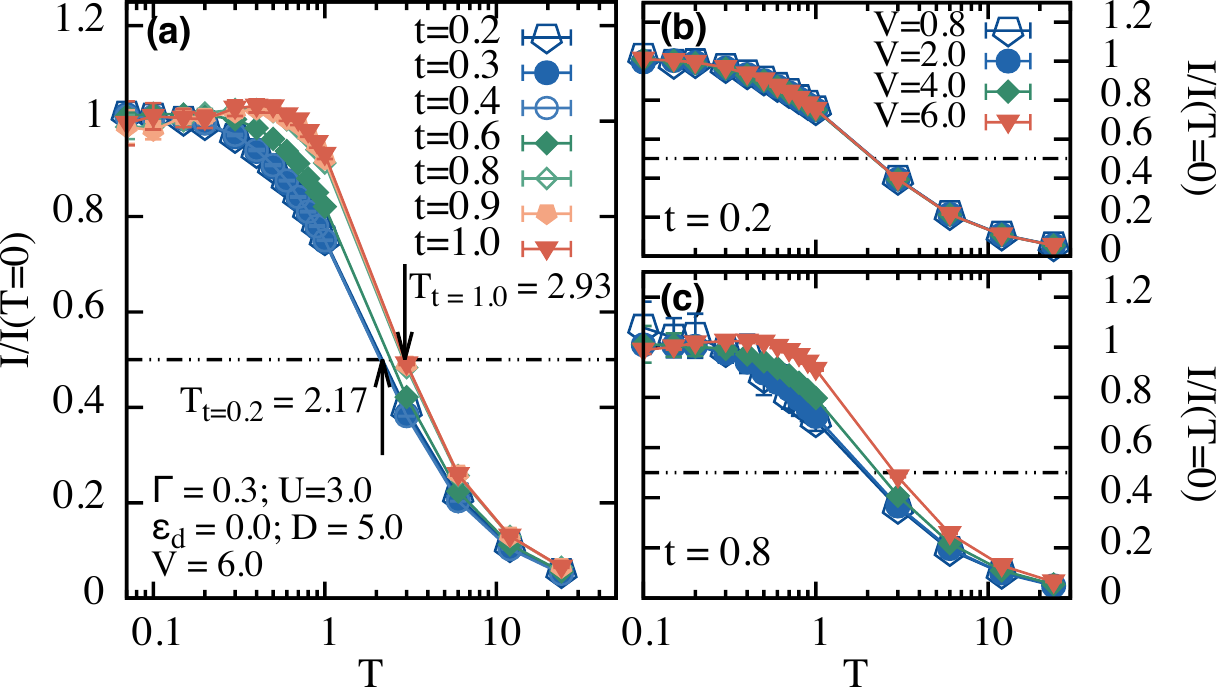}\vspace{-1.0em}
\caption{
(a) Normalized current $I/I(T=0)$ as a function of temperature at times $t=0.2, 0.4, 0.6,0.8, 1.0$ and $U=3, \varepsilon_d = 0.0, \Gamma = 0.3$, $D=5.0$ and voltage $V=6.0$. The dashed line indicates the value of $1/2$. (b) Normalized current $I/I(T=0)$ as a function of temperature at $t = 0.2$ and (c) $t = 0.8$ for different voltages.  }
\label{fig5}
\end{figure}

The temperature dependence of the current outside of the linear response regime is analyzed in Fig.~\ref{fig5}. Panel (a) shows $I/I(0)$ at $V = 6$ as a function of temperature $T$. Different traces show different transient times, between $0.2$ ($1$ ps) and $1$ ($4$ ps). Similarly to the linear response case (Fig. \ref{fig3}), the current exhibits the low temperature saturation at all times. The Kondo effect is therefore visible in transient dynamics for all voltages. Outside linear response, the temperature at which the current saturates is strongly time-dependent. We denote the characteristic saturation temperature as $T_{t}$ and define as before $I(T_t,t) = I(T=0,t) / 2$. At short times (panel (b)) $T_{t}$ is the Kondo temperature $T_K$ for all voltages. As time increases $T_{t}$ increases by $\approx 60\%$ and reaches its steady state value of $T_{t} \simeq 2.9$ for $V=6$ (c). Further increase of the voltage results in a non-monotonic temperature dependence of the current \cite{Reininghaus2014} (not shown).

\begin{figure}[ht!]
\includegraphics[width=\columnwidth]{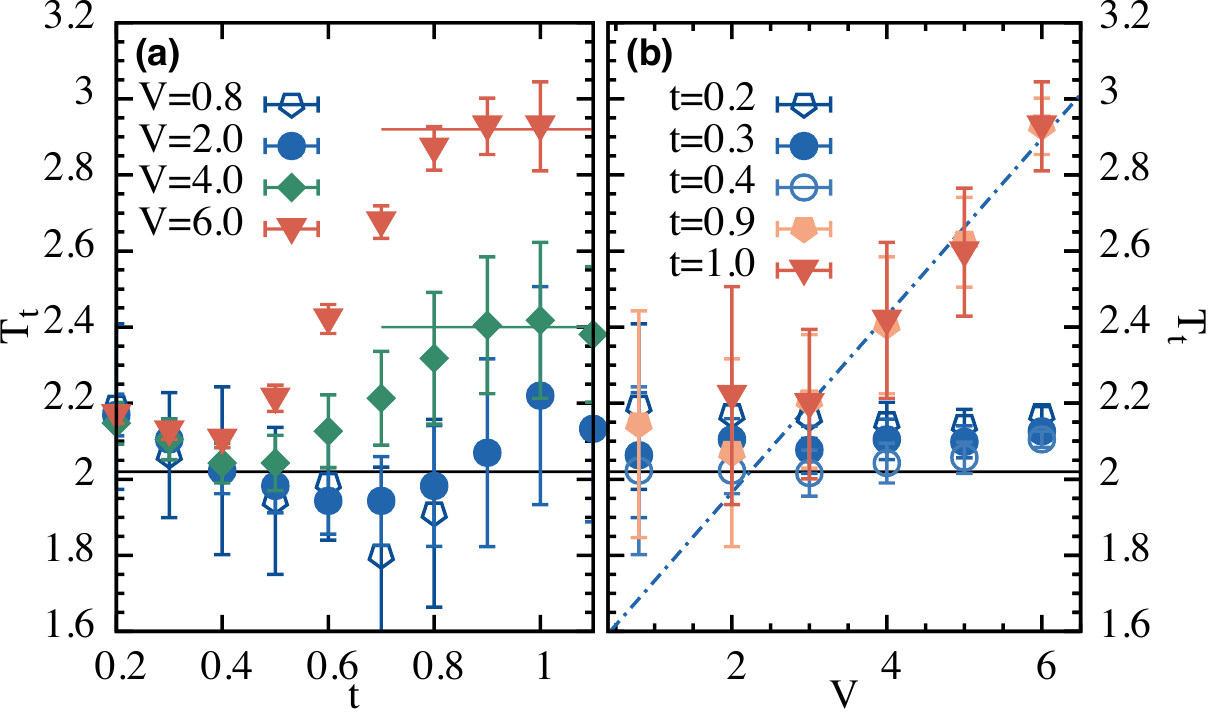}\vspace{-1.0em}
\caption{
(a) Current saturation temperature $T_{t}$ ($T$ at half of zero-temperature current value) as a function of time at different voltages at $U=3, \varepsilon_d = 0, \Gamma = 0.3$, $D=5$. Solid lines: estimates of the steady-state $T_{t}$. (b) $T_{t}$ as a function of voltage at a set of times. Solid line: linear response $T_{t} = T_K$, dashed line: linear fit at $V>2$. }
\label{fig6}
\end{figure}

 In Fig.~\ref{fig6} we show the time (a) and voltage (b) evolution of $T_{t}$.  For small voltages, there is no time-dependence of $T_{t}$. In contrast, as $V$ is increased up to $2U$, $T_{t}$ rapidly rises and equilibrates. Estimates of the steady-state $T_{t}$ as defined by the value at which the current reaches half of its low-$T$ value for $V=2,4,6$ are given by the horizontal lines. 
Fig.~\ref{fig6}b shows a separation of short- and long-time behavior: no change in $T_{t}$ is observed for short times $t \leq 0.4$ ($1.5$ ps), whereas $T_{t}$ in the steady state is found to be increasing with $V$ for $V > 2$ in agreement with experiments \cite{Kretinin2011} and predictions from the real time renormalization group \cite{Reininghaus2014}.

\emph{Conclusions}.~ We have described the transient dynamics of a quantum dot in the mixed valence regime following the instantaneous application of a bias voltage in a range of temperatures below and above the Kondo temperature $T_K$. We have observed the full dynamics of the system from equilibrium to steady state. At all times and all voltages below the lead bandwidth, the current saturates at low temperature, exhibiting Kondo behavior. In linear response the saturation temperature $T_t$ at which the current reaches half of its zero-temperature value is the Kondo temperature $T_K$. Outside of the linear response regime $T_t$ has a strong time dependence and connects the equilibrium Kondo temperature to the increased steady state value \cite{Reininghaus2014,Kretinin2011}. The temperature $T_t$ describes the interplay of non-equilibrium (applied voltage) and strong correlations in the system.

The results presented here are exact within the error-bars and the current and time-dependent $T_t$ are directly accessible in time-resolved experiments \cite{Nunes1993,Loth2010}. The dynamics of the mixed valence system is fast - the equilibration occurs on a picosecond scale. We believe that the same physics can be observed at much larger time scales by lowering the level spacing $\varepsilon_d$ and, consequently, entering the Kondo regime, thereby decreasing $T_K$ and exponentially increasing the relaxation time $\tau \propto \exp(-T_K)$ \cite{Gull2011}.

\begin{acknowledgments}
The authors acknowledge helpful discussions with Lucas Peeters, Guy Cohen, James P.~F. LeBlanc, H. Terletska, and Pedro Ribeiro, and thank DOE ER 46932 for financial support. We have used ALPSCore \cite{ALPS20,alpscore} and GFTools libraries \cite{gftools} for the programming development. This research used resources of the National Energy Research Scientific Computing Center, a DOE Office of Science User Facility supported by the Office of Science of the U.S. Department of Energy under Contract No. DE-AC02-05CH11231.
\end{acknowledgments}

\bibliographystyle{apsrev4-1}
\bibliography{paper_kondo}

\end{document}